\documentclass{article}

\usepackage[final, nonatbib]{neurips_2023}

\usepackage[utf8]{inputenc} 
\usepackage[T1]{fontenc}    
\usepackage{hyperref}       
\usepackage{url}            
\usepackage{booktabs}       
\usepackage{amsfonts}       
\usepackage{nicefrac}       
\usepackage{microtype}      
\usepackage{xcolor}         

\usepackage{microtype}
\usepackage{graphicx}
\usepackage{subfigure}
\usepackage{booktabs} 
\usepackage{multirow}

\usepackage{hyperref}
\usepackage{caption}

\usepackage{amsmath,amsthm,amssymb,amscd,url,enumerate, mathtools}
\usepackage[utf8]{inputenc}
\usepackage{wrapfig}
\usepackage{pdflscape}
\usepackage{tikz}
\usetikzlibrary{decorations.markings,arrows}
\usetikzlibrary{positioning,arrows,fit}

\usepackage{tikz-cd}
\usepackage{amsfonts}
\usepackage{amsmath}
\usepackage{graphicx}
\usepackage{amsthm}
\usepackage{mathrsfs}
\usepackage{dsfont}
\usepackage{enumitem}
\usepackage{array}
\usepackage{verbatim}
\newcolumntype{?}{!{\vrule width 1pt}}
\usepackage[capitalize,noabbrev]{cleveref}
\usepackage[font=footnotesize,labelfont=bf, labelsep=period, textfont=sl]{caption}

\newcommand{\para}[1]{{\vspace{2pt} \bf \noindent #1}}  
\newenvironment{packed_itemize}{
\begin{list}{\labelitemi}{\leftmargin=1em}
\setlength{\itemsep}{1pt}                                                           
\setlength{\parskip}{0pt}                                                                                 \setlength{\parsep}{0pt}                                                                                  \setlength{\headsep}{0pt}                                                                                 \setlength{\topskip}{0pt}                                                                                 \setlength{\topmargin}{0pt}                                                                               \setlength{\topsep}{0pt}                                                                                  \setlength{\partopsep}{0pt}                                                                               }{\end{list}}

\DeclarePairedDelimiter\ceil{\lceil}{\rceil}

\newcommand{\system}{{\textsc{Verde}}}
\newcommand{\salsys}{{\textsc{Salsa} \textsc{Verde}}}
\newcommand{\salsa}{{\textsc{Salsa}}}
\newcommand{\picante}{{\textsc{Picante}}}

\newcommand{\lwe}{\textsc{lwe}}
\newcommand{\fplll}{{\textit{fplll}}}
\newcommand{\nomod}{{\textbf{NoMod}}}

\newcommand{\rev}[1]{\textcolor{black}{#1}}

\definecolor{lightgreen}{HTML}{117733}
\definecolor{lred}{HTML}{FF0000}
\newcommand{\success}[1]{\textcolor{lightgreen}{\textbf{#1}}}
\newcommand{\fail}[1]{\textcolor{lred}{{\em #1}}}

\newcommand*\samethanksagain[1][\value{footnote}]{\footnotemark[#1]}

 \author{ {\bf Cathy Yuanchen Li}  \\ FAIR, Meta  \and
 {\bf Emily Wenger} \\ The University of Chicago \and
 {\bf Zeyuan Allen-Zhu}  \\ FAIR, Meta \and
 {\bf Francois Charton}\thanks{Co-senior authors} \\ FAIR, Meta \and 
 {\bf Kristin Lauter}\samethanksagain \\ FAIR, Meta}

\title{\Large \bf \salsys{}: a machine learning attack on Learning With Errors with sparse small secrets}

\date{\today}

\begin{document}

\maketitle

\thispagestyle{empty}

\begin{abstract}

Learning with Errors (\lwe{}) is a hard math problem 
used in post-quantum cryptography. 
Homomorphic Encryption (HE) schemes rely on the hardness of the \lwe{} problem for their security, and two \lwe{}-based cryptosystems were recently standardized by NIST for digital signatures and key exchange (KEM).  Thus, it is critical to continue assessing the security of \lwe{} and specific parameter choices. For example, HE uses secrets \rev{with small entries}, and the HE community has considered standardizing small sparse secrets to improve efficiency and functionality.  However, prior work, \salsa{} and \picante{}, 
showed that ML attacks can recover sparse binary secrets. Building on these, we propose \system{}, an improved ML attack that can recover sparse binary, ternary, and narrow Gaussian secrets. Using improved preprocessing and secret recovery techniques, \system{} can attack \lwe{} with larger dimensions ($n=512$) and smaller moduli ($\log_2 q=12$ for $n=256$), using less time and power. We propose novel architectures for scaling. Finally, we develop a theory that explains the success of ML \lwe{} attacks. 
\end{abstract}

\section{Introduction}
\label{sec:intro}

Language models have been successfully applied to numerous practical science problems in recent years. For example, transformers \cite{transformer17} have been used to solve problems in mathematics \cite{lample2019deep,davies2021}, theoretical physics \cite{dersy22}, chemistry \cite{schwaller19}, and biology \cite{rives19}. In this paper, we present an application of transformers to computer security: the cryptanalysis of Learning With Errors (\lwe{}) \cite{Reg05}, a hard math problem underpinning leading proposals for post-quantum public key cryptography.  

\textbf{Public-key cryptosystems} are the main solution for secure communication over the Internet. Public keys can be used to encode messages or verify digital signatures to or from a user with the corresponding private key. Security relies on the fact that recovering the private key from the public data requires solving a computationally hard math problem. Most currently deployed public-key systems are based on RSA \cite{rivest1978method}, which relies on the hardness of factoring large numbers into products of primes. Unfortunately, large-scale quantum computers will enable implementation of Shor's algorithm \cite{shor1994algorithms}, which can factor integers in quantum polynomial time and break such systems.
As a result, new hard problems are sought to serve as the basis of post-quantum public key cryptography (PQC). The US National Institute of Standards and Technology (NIST) ran a 5 year competition to define future PQC standards \cite{alagic2020status}, and standardized $4$ PQC systems in July 2022 \cite{nist2022finalists}. Two of these rely on special cases of the same hard (and Shor-free) problem: Learning with Errors (\lwe{}). 

\textbf{The Learning With Errors} problem (\lwe{}) assumes that it is hard to recover a secret vector $\textbf{s}$, given many \lwe{} samples $(\textbf a,b)$. In a \lwe{} sample, 
each $\textbf a$ is a random vector of $n$ integers modulo $q$ ($n$ is the dimension and $q$ is the modulus), and $b$ is a noisy modular inner product of $\textbf a$ and the secret key $\textbf s$\textemdash that is, $b=\textbf a \cdot \textbf s +e \mod q$, with the error $e$ drawn from a Gaussian distribution of width~$\sigma_e$ centered at $0$, $e \sim N(0, \sigma_e^2)$ and $\sigma_e \ll q$.  The hardness of \lwe{} is related to the hardness of well-known lattice problems such as the (approximate) Shortest Vector Problem (SVP). 

\textbf{Most classical attacks on \lwe{}} rely on lattice reduction \cite{LLL,BKZ,CN11_BKZ}. 
For example, given $m$ samples $(\mathbf{a_i}, b_i)$, 
create a matrix $A_{m\times n}$ whose $m$ rows are the vectors $\mathbf{a_i}$. The unique shortest vector problem (uSVP) attack recovers the secret $\textbf s$ by finding the shortest vector in a lattice constructed from $\bf b$, the columns of $A_{m\times n}$, and other parameters. 
The best known algorithms for solving SVP run in time exponential in the dimension~$n$.  Somewhat counter-intuitively, the smaller the modulus $q$, the harder the \lwe{} problem. 
Approximate solutions can be computed in polynomial time with LLL \cite{LLL}, but the approximation factor is exponentially bad (exponential in $n$).
We compare our results with uSVP attacks on \lwe{} in \S\ref{sec:Compare_uSVP}.

\textbf{Machine learning (ML) attacks on \lwe{}. \salsa{}}~\cite{wengersalsa}, the seminal ML attack on \lwe{}, 
uses a large collection of \lwe{} samples $\{(\textbf a,b)\}$ with the same secret $\mathbf s$ to train a transformer~\cite{transformer17} that predicts $b$ from $\textbf a$. 
\salsa{} presents methods for secret recovery via queries to the trained model, and observes that high model accuracy is not needed: secrets are recovered as soon as the transformer {\it starts} to learn (training loss drops).
\salsa{} is a proof of concept, recovering binary secrets with $3$ or $4$ nonzero bits for problems with dimension up to $128$, small instances of \lwe{} solvable via exhaustive search.

\textbf{\picante{}} builds on an observation from \salsa{} (\cite[Table 4]{wengersalsa}): transformers trained on \lwe{} samples~$\{(\mathbf a, b)\}$, with entries of $\mathbf a$ drawn from a restricted range instead of all of $[0,q)$, can recover binary secrets with larger Hamming weights $h$ (number of nonzero bits). So \picante{} introduces a data preprocessing phase during which the \lwe{} samples $(\mathbf a, b)$ are processed by BKZ, a lattice reduction algorithm, to obtain \lwe{} samples with the same secret but smaller coordinate  variance (and larger error, see~\S\ref{subsec:preprocessing}). 
In addition to training the transformer on the preprocessed samples, \picante{} reduces the number of \lwe{} samples required for the attack from $4$ million in \salsa{} to $4n$ (e.g.~$1400$ for~$n=350$), and improves secret recovery. Overall, \picante{} can recover binary secrets for dimensions up to $350$ and Hamming weight up to $60$. This is a considerable improvement over \salsa{}, faster than the uSVP attacks we compare against, and out of reach for exhaustive search.

\textbf{\picante{} has several limitations}. First, it only recovers sparse binary secrets, an important but limited subclass of \lwe{}. Homomorphic encryption (HE) may use binary secrets, but HE and other PQC schemes typically use ternary ($s_i \in \{-1,0,1\}$) or small ($|s_i|<k$, $k$ small) secrets. 
Second, \picante{}'s preprocessing is costly as dimension increases, making it difficult to scale \picante{} to dimensions larger than $350$. For $n=512$, $\log_2 q=45$, \picante{}'s preprocessing approach could not finish in a month with full parallelization. 
Third, \picante{} only experiments with large modulus $q$:  $\log_2 q=23$ for $n=256$, $\log_2 q= 27$ for $n=300$, and $\log_2 q= 32$ for $n=350$. Practical \lwe{}-based systems use small $q$: \textsc{Lizard} \cite{Lizard} recommends $\log_2 q=10$ for $n=608$ with sparse binary secrets, and the HE standard  \cite{HES} recommends $\log_2 q =25$ for $n=1024$ with ternary secrets.

\textbf{Our work, \salsys{}}, improves on \picante{} and makes the following contributions:
\begin{packed_itemize}
\item We introduce a {\bf two-bit distinguisher, a new secret recovery technique} for sparse binary, ternary and small Gaussian secrets. \system{} fully recovers binary and ternary secrets equally well (\S\ref{sec:secret_dist}). 
\item We improve data \textbf{preprocessing techniques}, making them forty times faster and $20\%$ more effective,
enabling recovery of binary, ternary and small Gaussian secrets for dimension $512$ (\S\ref{sec:dim_512}). 
\item We {\bf decrease the modulus $q$}, showing \system{} outperforming uSVP attacks (\S\ref{sec:smaller_q} and~\S\ref{sec:Compare_uSVP}).
\item We propose \nomod{}, a framework for understanding 
the success of ML-based \lwe{} attacks (\S\ref{sec:smaller_q}). 
\item We present a {\bf theoretical analysis} 
to show heuristically that successful secret recovery depends only on $\sqrt{h}$ and the standard deviation of the distribution of the \lwe{} data (\S\ref{sec:scaling_law}). 
\item We experiment with {\bf encoder-only models} and compare their success with seq2seq models (\S\ref{sec:model}). 
\end{packed_itemize}

\para{Key results.} Our main finding is that small, sparse \lwe{} secrets are weak. For dimension $n=256$, we can recover binary and ternary secrets with $10\%$ sparsity ($h/n$) for $\log_2 q= 20$, and with  $3\%$ sparsity when $\log_2q=12$ (Table~\ref{tab:highest_h_256}). 
For $n=512$, $\log_2 q=41$, we recover binary and ternary secrets with $\ge 11\%$ sparsity. Furthermore,
\system{} scales well to higher dimensions for small sparse secret recovery. Training \system{} models on $n=256/350/512$ problems takes only 1.5/1.6/2.5 hours per epoch, a small proportional increase compared to the increase in $n$. Also, we find that \system{} runs faster than uSVP attacks, at the expense of using more compute resources in parallel (see ~\ref{sec:Compare_uSVP}).

In Table~\ref{tab:highest_h_256}, we report the 
timings for successful secret recoveries, for binary/ternary/Gaussian secrets, with varying $n$ and $\log_q$. The table records the highest $h$ recovered for each column: for binary secrets, $h$ is the Hamming weight, and for ternary and Gaussian secrets, $h$ is the number of non-zero entries.  We record the amount of time needed for each stage of the attack, preprocessing (in hours/CPU, assuming full parallelization), model training (hours/epoch $\cdot$ (\# epochs)), and total attack time. For full parallelization, the number of CPU cores required is 4 million divided by $2n$. Source code and parameters to reproduce our main experiments are included in the supplementary material. The full code base will be open-sourced.

\begin{table}[h]
\small
  \centering
  \vspace{-0.3cm}
    \small\caption{\textbf{\system{} attack times (preprocessing, model training, and total), for dimension $n$ and $\log_2 q$. 
    \\ $h$ = \# non-zero entries in recovered secrets, $b$ = binary, $t$ = ternary, $g$ = Gaussian secret distributions.} 
    }
  \label{tab:highest_h_256}
  \vspace{1mm}
          \resizebox{0.99\textwidth}{!}{%
\begin{tabular}{lccccccccccccccc}
\toprule
$(n, \log_2 q)$    & \multicolumn{3}{c}{\hspace{-4mm} (256, 12)} &  \multicolumn{3}{c}{\hspace{-4mm} (256, 20)} & \multicolumn{3}{c}{\hspace{-4mm} (350, 21)} & \multicolumn{3}{c}{\hspace{-4mm} (350, 27)} & \multicolumn{3}{c}{\hspace{-1mm} (512, 41)} \\ \midrule
secret distribution & b  \hspace{-2mm}     & t  \hspace{-2mm}     & g  \hspace{2.5mm}     & b  \hspace{-2mm}     & t  \hspace{-2mm}     & g  \hspace{2.5mm}     & b  \hspace{-2mm}      & t  \hspace{-2mm}     & g  \hspace{2.5mm}    & b  \hspace{-2mm}      & t  \hspace{-2mm}     & g  \hspace{2.5mm}    & b  \hspace{-2mm}      & t  \hspace{-2mm}     & g     \\ 
highest $h$   & 8 \hspace{-2mm}       & 9   \hspace{-2mm}     & 5   \hspace{2.5mm}    & 33 \hspace{-2mm}      & 24    \hspace{-2mm}    & 7   \hspace{2.5mm}   & 12  \hspace{-2mm}     & 13  \hspace{-2mm}     & 5   \hspace{2.5mm}   & 36   \hspace{-2mm}     & 36   \hspace{-2mm}    & 10   \hspace{2.5mm}   & 63    \hspace{-2mm}    & 58  \hspace{-2mm}    & 16 \\ \midrule
preprocessing (hrs/CPU)   & 1.5 \hspace{-2mm}       & 1.5  \hspace{-2mm}     & 1.5  \hspace{2.5mm}    & 7.5  \hspace{-2mm}      & 7.5   \hspace{-2mm}    & 7.5  \hspace{2.5mm}   & 16  \hspace{-2mm}     & 16  \hspace{-2mm}     & 16   \hspace{2.5mm}   & 216   \hspace{-2mm}     & 216  \hspace{-2mm}    & 216  \hspace{2.5mm}   & 840    \hspace{-2mm}    & 840   \hspace{-2mm}    & 840 \\
training time (hrs) & 1.5 \hspace{-2mm}       & 3 \hspace{-2mm}     & 12  \hspace{2.5mm}    & 3  \hspace{-2mm}      & 7.5    \hspace{-2mm}    & 1.5   \hspace{2.5mm}   & 1.6 \hspace{-2mm}     & 25.6  \hspace{-2mm}     & 1.6   \hspace{2.5mm}   & 1.6 \hspace{-2mm}     & 17.6   \hspace{-2mm}    & 3.2   \hspace{2.5mm}   & 17.5    \hspace{-2mm}    & 27.5   \hspace{-2mm}    & 2.5 \\
total time (hrs)  & 3 \hspace{-2mm}       & 4.5   \hspace{-2mm}     & 13.5  \hspace{2.5mm}    & 10.5  \hspace{-2mm}      & 15    \hspace{-2mm}    & 9   \hspace{2.5mm}   & 17.6  \hspace{-2mm}     & 41.6 \hspace{-2mm}     & 17.6 \hspace{2.5mm}   &218   \hspace{-2mm}     & 234   \hspace{-2mm}    & 220   \hspace{2.5mm}   & 858   \hspace{-2mm}    & 868   \hspace{-2mm}    & 843 \\\bottomrule    
\end{tabular}
}
  \vspace{-2mm}
  \end{table}

\para{Scope of results.} Instances of cryptographic problems like LWE can be broadly categorized as easy (solvable via exhaustive search), medium-to-hard (requiring significant resources to solve), or standardized (believed secure). \system{} attacks {\em medium-to-hard} LWE problems (parameterized by dimension $n$, Hamming weight $h$, modulus $q$). \system{} does not attack toy problems (like \salsa{} did), nor does it attack the NIST standard directly. Rather, \system{} demonstrates successful attacks on medium-to-hard LWE problems using tools from AI, improving our understanding of the security of proposed LWE-based cryptosystems.

\section{\salsys{} Overview}
\label{sec:overview}
\vspace{-0.1cm}

In this section we describe the \salsys{} attack and relevant parts of its predecessor \picante{}.

\begin{figure}[h]
    \centering
    \includegraphics[width=0.9\textwidth]{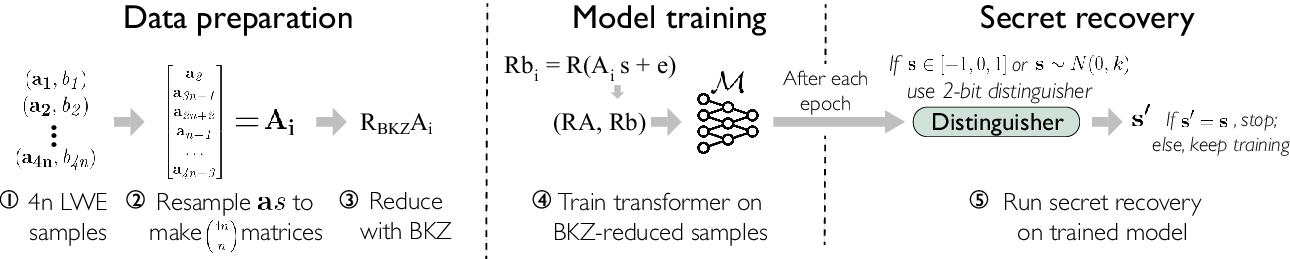}
    \caption{{\bf \em Overview of \system{}'s attack methodology}}
    \label{fig:overview_fig}
    \vspace{-0.2cm}
\end{figure}

\textbf{High-level overview.} Like \picante{}, \system{} starts with $4n$ \lwe{} samples with the same secret $\mathbf s$. In practice, this data would be eavesdropped. \system{} then proceeds in three stages: preprocessing, model training and secret recovery (see Figure~\ref{fig:overview_fig}). 
The preprocessing stage augments the $4n$ initial $(\textbf a,b)$ pairs to $2$ million, then runs lattice reduction to yield a training set of $4$ million samples with the same secret $\mathbf s$. 
The preprocessed data is used to train a transformer to predict $b$ from $\mathbf a$. After each training epoch (2 million \lwe{} samples), the model is queried to form a secret guess. The attack succeeds if the secret guess is correct, tested via statistical methods without knowledge of the actual secret. Otherwise, the model is trained for another epoch, and secret recovery is run again.

\para{Data preprocessing.}
\label{subsec:preprocessing}
\system{}'s preprocessing is similar to \picante{}'s, with several improvements. First, we create $n\times n$ matrices $\textbf A_i$ by sampling without replacement $n$ of the $4n$ original \lwe{} samples. Then, we apply lattice reduction to the matrices $\mathbf A_i$ to reduce the standard deviation of their entries (initially uniformly distributed over $[0,q)$). This process generates $2n$ preprocessed samples $(\mathbf a',b')$, with the same secret and is implemented in parallel to create a training set of $4$ million samples.

During lattice reduction, \picante{} applies BKZ (as implemented in \fplll{}~\cite{fplll}) to the $2n\times 2n$ matrix:
$\mathbf{ \Lambda}_i = \begin{bmatrix}
\omega\cdot\mathbf{I}_n & \mathbf{A}_i \\
0 & q\cdot \mathbf{I}_n
\end{bmatrix}$.
BKZ finds a linear transformation $\begin{bmatrix} \mathbf{R}_i & \mathbf{C}_i \end{bmatrix}$ such that the norms of the $2n$ rows of
$\begin{bmatrix} \mathbf{R}_i & \mathbf{C}_i \end{bmatrix} \mathbf{\Lambda}_i
= \begin{bmatrix} \omega \cdot\mathbf{R}_i & \mathbf{R}_i\mathbf{A}_i+q\cdot\mathbf{C}_i \end{bmatrix}$ are small.
Applying $\mathbf{R}_i$ to $\mathbf A_i$ and $\mathbf b_i$, \picante{} generates $2n$ reduced \lwe{} pairs $(\mathbf R_i \mathbf A_i,\mathbf R_i \mathbf b_i)$ (modulo $q$). \system{} instead rearranges the rows of $\mathbf \Lambda_i$ and applies lattice reduction to $ \mathbf{\Lambda'}_i =
 \begin{bmatrix}
0 & q\cdot \mathbf{I}_n \\
\omega\cdot\mathbf{I}_n & \mathbf{A}_i 
\end{bmatrix}$. 
This reduces the number of operations needed for lattice reduction and allows BKZ to run with lower floating point precision. These two improvements cut the preprocessing time significantly.

They also result in smaller $\mathbf R_i$ norms (smaller error in $\mathbf R_i \mathbf b_i$), which can be leveraged to further reduce the norm of $\mathbf{R_iA_i}+q\cdot\mathbf{C_i}$ by using a lower $\omega$ parameter ($\omega = 10$) than \picante{} ($\omega = 15$). 
Finally, \system{} replaces BKZ with two interleaved algorithms (BKZ~2.0~\cite{CN11_BKZ} and the efficient reduction technique introduced in~\cite{charton2023efficient}), adaptively increases the blocksize and precision as reduction progresses, and introduces a stopping condition. 
For instance, for $n=256$ and $\log_2 q=20$, 
\system{}'s preprocessing is $45\times$ faster, while improving the quality of the reduction by $20\%$ (Table~\ref{tab:preprocess}). 

\begin{table}[h!]
\small
  \centering
\vspace{-3mm}

    \small\caption{\textbf{Impact of successive improvements to preprocessing}. $n=256$, $\log_2 q = 20$. Reduction factor of the standard deviation of entries of $\mathbf A_i$: lower is better. Time: \# of hours to preprocess one matrix on one CPU.}
    \label{tab:preprocess}
    \vspace{1mm}
        \resizebox{0.99\textwidth}{!}{%
\begin{tabular}{lcc}
\toprule
{\bf preprocessing technique} & {\bf reduction factor} & {\bf time (hrs/CPU)} \\ \midrule
\picante{}        & 0.53   & 338  \\
\hspace{1.2cm}   + reordered $\mathbf \Lambda _i$ rows     & 0.47   & 136  \\
\hspace{1.2cm}    + reduced floating point precision    & 0.47   & 24   \\
\hspace{1.2cm} + reduced parameter $\omega$ from 15 to 10      & 0.43   & 38   \\
\system{} (+ interleaved reduction, adaptive blocksize, early stopping)     & {\bf 0.43}   & {\bf 7.5}  \\ \bottomrule
\end{tabular}
}

\vspace{-1mm}
\end{table}

Quality of reduction is measured by the standard deviation of the entries of $\mathbf A_i$. 
Preprocessing time in Table~\ref{tab:preprocess} is the hours needed to process one matrix on a single CPU. We list the time for all $n,q$ attempted in Tables~\ref{tab:512_results} and \ref{tab:gaussian_verde}. For each dimension $n$ and modulus $q$, we process 2 million$/n$ matrices in parallel across hundreds of CPUs, see \S\ref{subsec:params} for details.  
Better preprocessing allows \system{} to scale to larger dimensions and retrieve secrets with larger $h$.

\para{Transformer training.}
\label{subsec:training}
The $4$ million reduced \lwe{} pairs are used to train a transformer to predict~$b$ from~$\mathbf a$. The values $b$ and the coordinates of $\mathbf a$ are integers in $[0,q)$. They are represented in base $B = \ceil{q/8}$ (for $\log_2 q > 30$, $B=\ceil{q/16}$) and encoded as sequences of two tokens over a vocabulary of $2,000$ 
(see \S\ref{subsec:params} for a discussion of these choices). Model training is framed as a translation task, from a sequence of $2n$ tokens representing $\mathbf a$ to a sequence of $2$ tokens representing $b$ (see \cite{lample2019deep, charton2021linear} for similar uses of transformers for mathematical calculations). The model is trained to minimize the cross-entropy between model prediction and the sequence of tokens representing $b$, using the Adam optimizer with warmup \cite{kingma2014adam} and a learning rate of $10^{-5}$. For $n=256,350$ and $512$, each epoch uses $2$ million \lwe{} samples and runs for $1.5, 1.6$, or $2.5$ hours. 
Time/epoch doesn't vary with $q$ or secret type.
Our models train on one NVIDIA V100 32GB GPU  and often succeed in the first epoch for low $h$. Number of epochs required are included in many tables throughout, including 
Tables ~\ref{tab:ternary_n256_q20} and \ref{tab:gaussian_n350_q27}. 

\system{} uses the same architecture as \picante{}: a sequence to sequence (seq2seq) transformer \cite{transformer17}, with a one-layer encoder (dimension $1024$, $4$ attention heads), and a 9-layer decoder (dimension $512$, $4$ heads). The last $8$ layers of the decoder are shared (i.e. they form a Universal Transformer~\cite{dehghani2018universal}). Iteration through shared loops is controlled by the copy-gate mechanism introduced in~\cite{csordas2021neural}. 

Seq2seq models allow output sequences to be longer than inputs, a useful feature for machine translation but not necessary in our setting. For comparison, we also implement a simpler \emph{encoder-only} transformer, that is 4-layer BERT-like (dimension $512$, $4$ heads), together with rotary word embeddings (analogous to the rotary position embeddings~\cite{rope21}) to account for the modular nature of the problem. 
On top of this, we also add an earth mover's distance (EMD) auxiliary objective.
We compare this model's performance to that of the seq2seq model (\S\ref{sec:model}).

\para{Secret recovery.}
\label{subsec:secret_recovery}
Secret recovery runs after each epoch ($2$ million \lwe{} samples). \picante{} used three recovery methods: direct recovery, cross-attention and distinguisher. Direct recovery struggles as dimension increases, because it relies on accurate model evaluations at special $\textbf a$ values which are out of distribution for the training set. 
Cross-attention is incompatible with encoder-only architectures and  is consistently outperformed by the distinguisher on \system{}-preprocessed data. Thus, \system{} only uses the distinguisher, which works as follows: for any test vector $\textbf a_{\text{test}}$ and random $K$, if the $i$-th entry of the secret $s_i$ is zero, then $(\mathbf a_{\text{test}}+K\mathbf e_i) \cdot \mathbf s = \mathbf a_{\text{test}}\cdot \mathbf s$ (where $\mathbf e_i$ is the $i$-th standard basis vector). Therefore, for each $i$, the distinguisher computes the difference between model predictions on $\bf a_{\text{test}}$ and $\mathbf a_{\text{test}} + K\mathbf e_i$ for different $\mathbf a_{\text{test}}$. If differences are small, the corresponding $s_i$ is likely zero. For ternary and Gaussian secrets, the distinguisher is modified (see \S\ref{sec:secret_dist}).

For successful secret recovery, the trained model must generalize well to $\{\mathbf a_{\text{test}}\}$, the vectors used for testing. In \picante{}, the distinguisher runs on random $\mathbf a_{\text{test}}$, with coordinates uniform in $[0,q)$. However, the model is trained on preprocessed $\textbf a_{\text{train}}$, with a non-uniform coordinate distribution. So \picante{}'s distinguisher recovery requires that the model generalize outside its training distribution. This is known to be a difficult ML task.
Instead, \system{} runs the distinguisher on a held-out subset of $128$ preprocessed vectors $\mathbf a_{\text{test}}$. Since the test vectors have the same distribution as the training set, the trained model only needs to generalize in-distribution, a much easier task. 
This change in $\textbf a_{\text{test}}$ improves the performance of the distinguisher (see Table~\ref{tab:distinguisher_bkz} in~\S\ref{subsec:dist_bkz}).

In practice, for each secret coordinate, the distinguisher computes the sum of the absolute difference between model predictions at $\mathbf a_{\text{test}}$ and $\mathbf a_{\text{test}}+K\mathbf e_i$, for $128$ vectors $\mathbf a_{\text{test}}$ and a random $K \in(0.3q,0.7q)$ for each $\mathbf a_{\text{test}}$. The model makes a secret prediction $\textbf s'$ by setting the bits to $1$ on the $h$ coordinates with the largest sums, and verifies $\textbf s'$ by computing $\mathbf a \cdot \mathbf s' - b$ on the original $4n$ \lwe{} samples. If the secret is correctly predicted, this quantity should always be small. The statistical properties of this test are discussed in section A.2 of \cite{li2023salsa}.  
Having discussed \system{}'s background and described its methodology, we now present \system{}'s key results, as summarized in \S\ref{sec:intro}.
We begin with results on ternary and narrow Gaussian secrets.

\vspace{-0.1cm}
\section{Secret Distributions} 
\label{sec:secret_dist}
\vspace{-0.1cm}

One of \system{}'s major contributions is a method to recover secrets from sparse ternary and narrow Gaussian distributions, both of which are being considered for use in real-world cryptosystems~\cite{HES}. Prior ML-based LWE attacks (\picante{} and \salsa{}) only recovered sparse binary secrets. 
Here, we describe \system{}'s method for recovering these more general secret distributions, and its performance on ternary and small Gaussian secrets for fixed dimension $n=256$ and $\log_2q = 20$. Throughout this paper, $h$ denotes the number of nonzero entries in a secret, which is equal to the Hamming weight in the binary case. We define secret {\it sparsity} as the percentage of nonzero secret entries $h/n$.

\para{Recovering ternary and small Gaussian secrets.} In lattice-based cryptography, ternary secrets are vectors $\textbf s$ of dimension $n$, with entries equal to $0$, $1$, or $-1$ (with equal probability of $1$ and $-1$ in sparse secrets).  Gaussian secrets are vectors $\textbf s$ of dimension $n$, with entries drawn from a Gaussian distribution with small standard deviation $\sigma$. In this paper, we use $\sigma=3$. The Homomorphic Encryption Standard~\cite{HES} includes secure parameter choices for Gaussian secrets with $\sigma=3.2$.

Ternary and Gaussian secrets introduce new challenges for ML-based attacks on \lwe{}. The recovery methods in \picante{} distinguish zero from nonzero bits. In the binary case, this produces one guess~$\textbf s_{\text{guess}}$, which can be verified by checking that $b-\textbf a \cdot \textbf s_{\text{guess}}$ is small. In the ternary case, \picante{} would produce $2^h$ secret guesses (about $12^h$ guesses for Gaussian secrets), due to the additional $-1$ entries, and verification becomes very expensive as $h$ increases. 
\system{} recovers ternary secrets using a two-step approach: first, \emph{partial recovery} distinguishes zero from nonzero entries, then \emph{full recovery} guesses the sign of the nonzero bits.

{\em Partial recovery.} To identify nonzero bits, \system{} uses the binary secret distinguisher from \S\ref{subsec:secret_recovery} (after~\cite[Section 4.3]{li2023salsa}). For each secret bit, it computes a score from a sample of reduced \lwe{} pairs. The $h$ bits with the highest scores are candidate nonzero bits. \system{} assumes $h$ is not known and runs the next step for all reasonable possible values of $h$, e.g. from $1$ to $n/20$. Partial recovery alone is a major contribution, as we are not aware of existing attacks that can identify nonzero secret bits. 

{\em Full recovery (ternary secrets).} To determine whether nonzero bits of a ternary secret are $1$ or $-1$, \system{} introduces a novel two-bit distinguisher, leveraging the following observation. If two nonzero secret bits $s_i$ and $s_j$ are equal, then for any $\textbf a$, exchanging the coordinates $a_i$ and $a_j$ will result in the same $b$, and corresponding model predictions will be close. Otherwise, model predictions will be different (if $s_i \neq s_j$). Similarly, if $s_i = s_j$, for any $\mathbf a$ and $c \neq 0$, changing $a_i \rightarrow a_i + c$ and $a_j \rightarrow a_j - c$ yields the same $b$, and close model predictions. The two-bit distinguisher uses these techniques to compare each nonzero bit with all others, therefore defining two classes of nonzero bits. Letting one class of bits be $1$ or $-1$, \system{} produces two secret guesses to be verified.  

{\em Full recovery (Gaussian secrets).} At present, we implemented full recovery only for binary and ternary secrets. However, full recovery of small Gaussian secrets is possible via the following adaptation of the two-bit distinguisher. The two-bit distinguisher groups nonzero secret bits into $k$ classes believed to have the same value. In our case, the nonzero bits follow a Gaussian distribution with $\sigma=3$, so we may safely assume that all non-zero secret bits are in $[-9, 9]$ (within $3$ standard deviations) -- i.e. $k=18$. Since the secret is Gaussian, we expect the largest classes to correspond to the values $-1$ and $1$, followed by $-2$ and $2$, and so on. Therefore, 
 we can intelligently assign values to classes based on class size, and test the corresponding $2^{k/2}=512$ secrets.  We leave implementation of this as future work and report the performance of partial Gaussian secret recovery, using knowledge of $\mathbf{s}$ to validate correctness.

\begin{table}[h]
\small
  \centering

\vspace{-2mm}
\small\caption{\textbf{Partial and full ternary secret recovery. } $n=256, \log_2 q=20$. Epoch when secret is recovered.}
  \label{tab:ternary_n256_q20}
  \vspace{1mm}
    \resizebox{0.99\textwidth}{!}{%
\begin{tabular}{llllllllll}
\toprule
{ $h$}                                                                  & 5               & 10          & 15          & 20   & 21      & 22   & 23   & 24    & 25   \\ \midrule
partial recovery                                                                                                            & 8/10            & 6/10        & 6/10        & 2/10 & 3/10    & 2/10 & 1/10 & 3/10  & 1/10 \\
{training epoch}                                                                                                           & 0,0,0,0,0,0,1,7 & 0,0,1,1,1,1 & 0,0,0,1,2,8 & 0,2  & 0,1,3   & 2,5  & 9    & 0,4,7 & 2    \\ \midrule
full recovery                                                                                                            & 8/10            & 6/10        & 5/10        & 1/10 & 3/10    & 2/10 & 0/10 & 1/10  & 0/10 \\
{training epoch}                                                                                                        & 0,0,0,0,0,0,1,7 & 1,2,2,6,7,8 & 1,1,6,9,11  & 5    & 4,10,17 & 8,22 &      & 5,12  &  \\ \bottomrule    
\end{tabular}
}
  \vspace{-0.3cm}
\end{table}

\begin{table}[h]

\small
  \centering
\small\caption{\textbf{Partial Gaussian secret recovery.} $n=350, \log_2 q=27$. Epoch when secret is recovered. }
  \label{tab:gaussian_n350_q27}
    \vspace{1mm}
    \resizebox{0.99\textwidth}{!}{%
\begin{tabular}{lllllllll}
\toprule
 $h$ & 4               & 5               & 6              & 7          & 8    & 9    & 10       \\ \midrule
partial recovery                  & 8/10            & 8/10            & 7/10           & 5/10       & 2/10 & 2/10 & 4/10     \\
{training epoch}                & 0,0,0,0,1,1,1,1 & 0,0,0,1,2,3,3,7 & 0,0,1,2,2,5,10 & 2,2,3,7,10 & 0,2  & 1,9  & 1,5,6,12 \\ \bottomrule
\end{tabular}
}
    \vspace{0.1cm}
\end{table}

\begin{wrapfigure}{r}{0.4\textwidth}
\vspace{-4mm}
  \begin{center}
    \includegraphics[width=0.38\textwidth]{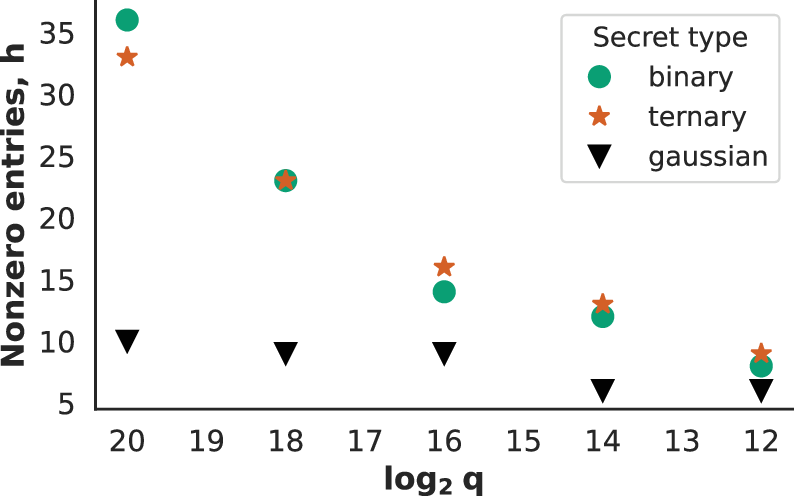}
  \end{center}
  \vspace{-3mm}
\caption{{\bf \em Best $h$ recovered vs. $\log_2q$ and secret distribution}. $n=256$.}
    \label{fig:secret_types}
\vspace{-5mm}
\end{wrapfigure}
\textbf{\system{}'s performance across secret distributions.} \system{} recovers {\em ternary secrets} with sparsity up to $10\%$, with comparable performance on binary secrets. 
Table~\ref{tab:ternary_n256_q20} provides details on partial and full ternary secret recovery for $n=256$ and $\log_2 q = 20$ and $h =5--25$. For low values of $h$ ($h<20$), ternary secrets are partially recovered early during training (i.e. mostly in epoch $0$ or $1$, during the first pass on preprocessed data), and usually fully recovered in the same epoch or shortly after. As $h$ increases, more training is required for recovery, and the delay between partial and full recovery increases.

For {\em small Gaussian secrets}, \system{} only implements partial recovery (recovery of the nonzero bits). Table~\ref{tab:gaussian_n350_q27} presents results for $n=350$ and $\log_2 q = 27$. Recovered $h$ are lower than in the binary case, and models require less training for low $h$.

Figure~\ref{fig:secret_types} compares \system{}'s performance across secret distributions for problems with $n=256$ and different moduli $q$. For each setting, we run $100$ recovery experiments and report the highest $h$ secret recovered in those attempts. Recovery is comparable for binary and ternary secrets. Small Gaussian secrets are significantly harder.

\section{Large Dimension} 
\label{sec:dim_512}
The hardness of \lwe{} increases as $n$ grows.
\picante{} recovered sparse binary secrets for dimension up  to $n=350$. \system{} pushes this limit to $n=512$, for sparse binary, ternary, and narrow Gaussian secrets. 
For $n=512$ and $\log_2 q=41$, \system{} recovers binary and ternary secrets with sparsity up to $0.12$ (highest $h = 63,60$) and Gaussian secrets with $h$ up to $16$. The longer \system{}'s preprocessing step runs, the higher $h$ secrets it can recover.
Recall that real-world schemes like LIZARD operate in dimension $n=608$~\cite{Lizard} and HE in dimension $n=1024$~\cite{HES}. Those systems use significantly smaller $q$ where \lwe{} is harder: 
$\log_2 q=10$ and $27$, respectively, compared to \system{}'s $\log_2 q = 41$.

\para{\system{}'s performance in large dimension.} Table~\ref{tab:512_results} shows \system{}'s performance for dimension $n=512$, after $7$ to $35$ days of preprocessing, on the same set of matrices $\mathbf A_i$. For each preprocessing time, we measure the quality of lattice reduction via a ``reduction factor,'' computed by taking the ratio of the standard deviation of reduced $\mathbf A_i$ entries to the standard deviation of a random matrix with uniform coefficients $0 \leq a_i < q$, i.e. 
$\frac{\text{stddev}(\mathbf A_{\text{bkz}})}{\text{stddev}(\mathbf A_{\text{rand}})}$, with 
$\text{stddev}(\mathbf A_{\text{rand}}) = \frac{q}{\sqrt{12}}$. This metric, used in \picante{} for selecting BKZ parameters, is discussed in \S~\ref{sec:scaling_law}. 
As Table~\ref{tab:512_results} demonstrates, the maximum recoverable $h$ is strongly correlated to the quality of lattice reduction.

\begin{table}[h!]
\small
  \centering

  \vspace{-0.5cm}
\small\caption{\textbf{Data preprocessing vs performance}. $n=512$, $\log_2 q=41$. Highest values of $h$ recovered, for different reduction factors (lower factor $=$ better reduction).}
\label{tab:512_results}
\vspace{1mm}
    \resizebox{0.99\textwidth}{!}{%
\begin{tabular}{cclll}
\toprule
\textbf{preprocess time} & \textbf{reduction factor} &  \textbf{binary $h$}   & \textbf{ternary $h$}    & \textbf{Gaussian $h$}     \\ \midrule
7 days  & 0.519 & 16,17,17,20       & 17,20,20,21,21    & 8,8,8,8,10,10,13           \\
10 days & 0.469 & 21,22,23,28                   & 22,24,24,27,27,29       & 11,11,11,12,12,12          \\
14 days & 0.423 & 32,32,34,34,35,40             & 32,34,34,35,35          & 11,11,11,12,12,12          \\
20 days & 0.380  & 35,35,36,41,49                & 35,35,37,45,46          & 13,13,13,16 \\
28 days & 0.343 & 40,43,45,47,50,51,55          & 40,41,41,44,45,48,48,53 & 13,13,14,16,16             \\
35 days & 0.323 & 48,48,49,52,57,59,63         &   45,46,50,55,58,60       & 14,16                            \\ \bottomrule
\end{tabular}
}
\vspace{-0.2cm}
\end{table}

\para{Preprocessing adjustments for large $n$.} Scaling up to $n=512$ requires a number of adjustments to our preprocessing methodology. For $n=512$, the first loops in BKZ 2.0 are very slow. To avoid this, we use BKZ with smaller blocksizes than those used for $n=256$ and $350$ (see \S\ref{subsec:params}). 
Also, lattice reduction is significantly slower for larger matrices. To mitigate this, for $n=512$, we use $448$ \lwe{} samples (instead of $512$) when generating $\textbf A_i$ for lattice reduction, therefore reducing the matrix size from $1024\times 1024$ to $960\times 960$. Experimentally, we observe that using slightly fewer samples did not negatively impact our reduction factor or attack performance. 
\section{Small Modulus}
\label{sec:smaller_q}

\begin{figure*}[h]
\vspace{-0.3cm}
    \begin{center}
           \begin{minipage}[c]{0.48\linewidth}
           \vspace{0.35cm}
            \centering
\includegraphics[width=0.68\textwidth, height=3.2cm]{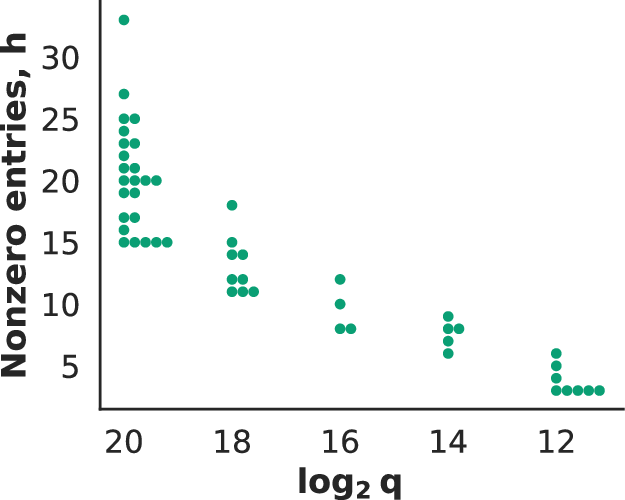}
            \caption{\textbf{$h$ of recovered secrets vs. $\log_2 q$,} $n=256$. 10 random binary secrets attempted for each $h$. One green dot represents a successful recovery.}
            \label{fig:q_vs_h_binary}
        \end{minipage}\hfill
        \begin{minipage}[c]{0.48\linewidth}
        \centering
                  \vspace{2mm}
            \small\captionof{table}{\textbf{Highest $h$ recovered, $n=256$, $350$}. Secret distributions are b = binary, t = ternary, g = Gaussian. }
        \label{tab:small_q}
        \resizebox{0.72\textwidth}{!}{%
            \begin{tabular}{ccccc}
            \toprule
            \multirow{2}{*}{$n, \log_2 q$} & \multirow{2}{*}{\begin{tabular}[c]{@{}c@{}}{\bf reduction}\\ {\bf factor}\end{tabular}} & \multicolumn{3}{c}{{\bf recovered} $h$} \\
             &  & b & t & g \\ \midrule
            256, 20 & 0.43 & 33 & 24 & 7 \\ 
            256, 18 & 0.53 & 18 & 19 & 7 \\
            256, 16 & 0.63 & 12 & 12 & 6 \\
            256, 14 & 0.71 & 9 & 9 & 6 \\
            256, 12 & 0.77 & 6 & 6 & 5 \\ \midrule
            350, 27 & 0.38 & 36 & 36 & 10 \\ 
            350, 21 & 0.61 & 12 & 13 & 5 \\ \bottomrule
            \end{tabular}
                }
        \end{minipage}
    \end{center}
    \vspace{-0.3cm}
\end{figure*}

To define real-world parameters for lattice-based cryptography, standardization committees and communities (e.g. \cite{HES,nist2022finalists}) 
select a small enough modulus $q$ (for fixed dimension $n$), so that all known attacks are (heuristically) predicted to run in time at least $2^{128}$, therefore attaining the U.S. government minimum $128$-bit security level. For classical lattice reduction attacks, the smaller the modulus, the more difficult the attack. 
This is because lattice reduction algorithms such as LLL and BKZ attempt to compute short vectors in Euclidean space, but cryptosystems operate modulo $q$. Smaller moduli result in smaller lattice volumes, meaning shorter target vectors are required to break the system. In our ML approach, we also observe that \system{} is less likely to succeed when $q$ is smaller (see Table \ref{tab:small_q}). Nevertheless, \system{} outperforms the uSVP attack (\S\ref{sec:Compare_uSVP}).

\para{\picante{} vs. \system{}'s performance on small $q$}. \picante{} attacks larger moduli: it can recover binary secrets with $h=31$ for $\log_2 q=23$ and $n=256$, and $h=60$ for $\log_2 q=32$ and $n=350$. 
Table~\ref{tab:small_q} presents \system{}'s highest recovered $h$ (in 10 random attempts) for dimensions $256$ and $350$ and different $q$ 
for binary, ternary, and narrow Gaussian secrets. First, note that
\system{} recovers binary secrets with $h=33$ for $n=256$ and $\log_2 q = 20$, but also, \system{} succeeds for much smaller $q$, as small as 
 $\log_2 q=12$, a near-real-world parameter setting,
 demonstrating that \system{} significantly outperforms \picante{}. However, for smaller $q$, \system{} recovers only secrets with smaller $h$. 
As with classical attacks, the likely culprit is lattice reduction: the reduction factor after preprocessing  is $0.71$ for  $\log_2 q = 12$ 
versus $0.43$ for $\log_2 q = 20$, and $h$ for recovered binary secrets drops from $33$ to $6$. Section ~\ref{sec:scaling_law} provides a theoretical explanation of this phenomenon.

Figure~\ref{fig:q_vs_h_binary} visualizes \system{}'s success rates for $n=256$ with binary secrets. For every value of $q$ and $h$, we run \system{} on $10$ binary secrets, using the same preprocessed data.
\system{}'s success rate decreases as $h$ increases. 
Attempting $10$ random binary secrets for $n=256, \log_2q=12$, \system{} recovers secrets for up to $h=6$ but not $h=7,8$ (Table \ref{tab:small_q}). However, with more attempts, \system{} recovers $1/100$ binary secrets for $h=8$.
Experiments with different random seeds (see \S\ref{subsec:seeds}) suggest that model initialization alone is not responsible for how success rate trends with $h$. 

\textbf{Explaining success for smaller $q$ via \nomod{}.} As Table~\ref{tab:small_q} indicates, for given $n$ and $h$, secrets are harder to recover for smaller $q$. This suggests that the modular operations in the computation of $b$ from $\textbf a$ might account for the difficulty. To investigate this, we evaluate, for a given known binary secret $\mathbf s$, the percentage of samples where computing $b$ did not require the modular operation, for the $4$ million samples in our training set. More precisely, we represent 
the mod $q$ coordinates of $\textbf a$ and $b$ in the interval $(-q/2,q/2)$, and compute $x = \mathbf{a} \cdot \mathbf{s} - b$
without modular operations. If $b$ was computed without modular operations, then $x$ is equal to the error for that sample, which is small.  Otherwise $x$ is equal to the error plus a multiple of $q$.
For each $(\textbf a,b)$, if $|x|<q/2$, then no modular operation was performed. 
We define \nomod{} to be the percentage of such $x$ in the training set. 

\begin{figure*}[t]
    \begin{center}
\begin{minipage}[c]{0.32\linewidth}
\centering
    \includegraphics[width=0.9\textwidth]{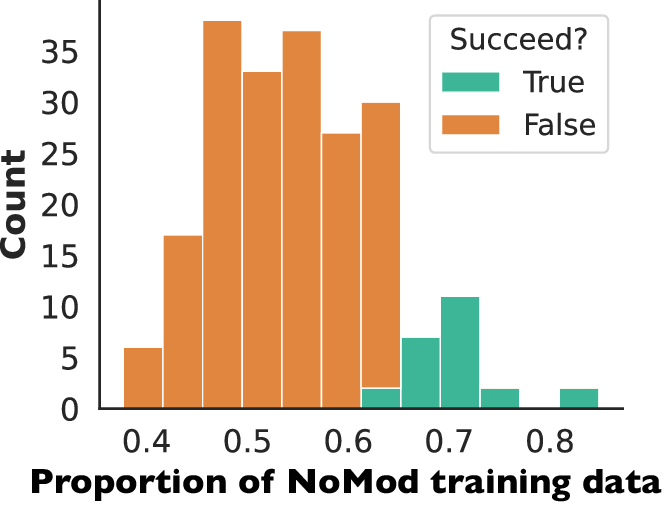}
    \vspace{-0.1cm}
    \caption{{\bf \em Effect of \nomod{} data on secret recovery for $n=256$, binary secrets.} Count = \# of experiments.}
    \label{fig:nomod_hist}
\end{minipage} \hspace{2mm}
        \begin{minipage}[c]{0.65\linewidth}
  \centering
      \small\captionof{table}{\textbf{\nomod{} $\%$ before/after permuting the columns of $A$}. $n=256, \log_2 q = 14$, binary secrets. Each column is a random secret for each $h = 6, 7, 8$. 
      Entries are the \nomod{} $\%$. 
    \success{Green} $=$ secret recovered; \fail{\em red} or black $=$ failure.}
    \label{tab:permute_cols}
\resizebox{0.92\textwidth}{!}{
\begin{tabular}{llllllllllll}
\toprule
$\bf h$ & {\bf method} & \multicolumn{10}{c}{\bf \nomod{} percentages and recovery success} \\ \midrule
\multirow{2}{*}{$6$}  & original & 56 & 61 & 60 & 61 & 56 & \fail{52} & 67 & 67 & \success{76} & 67  \\
                    & permuted & 57 & 67 & 56 & 67 & 62 & \success{67} & 60 & 60 & \fail{52} & 56 \\ \midrule
\multirow{2}{*}{$7$}  & original & \fail{60} & 60 & 52 & \fail{49} & 60 & \success{75} & 55 & \fail{60} & 55 & 56  \\
                    & permuted & \success{67} & 52 & 60 & \success{75} & 60 & \success{73} & 60 & \success{71} & 59 & 66 \\ \midrule
 \multirow{2}{*}{$8$}  & original & 60 & \success{74} & \success{74} & 66 & 60 & 66 & 55 & 60 & 63 & 55  \\ 
                    & permuted & 55 & \fail{55} & \fail{60} & 52 & 55 & 60 & 55 & 49 & 60 & 60 \\ \bottomrule
\end{tabular}
}
\end{minipage}\hfill
\end{center}
\vspace{-0.5cm}
\end{figure*}

Figure~\ref{fig:nomod_hist} shows the distribution of secret recoveries, for varying \nomod{}, for 210 experiments with dimension $n=256$ and $\log_2 q=20$. Clearly, recovery occurs when \nomod{} exceeds a threshold value, empirically observed to be $67\%$. These results confirm our hypothesis that modular arithmetic is the major cause of failure to recover secrets, and help explain the observation (\cite[\S6.4]{li2023salsa} and \S\ref{subsec:seeds}) that some secrets are never recovered, no matter how many model initializations are tried. For a given secret, \nomod{} is a property of the training data, not of model initialization, and multiple initializations only seem to help when \nomod{} is close to the recovery threshold.

\textbf{A trick for improving attack success.}  The \nomod{} percentage can only be calculated if the secret is known, so it cannot aid real-world attack settings. However, our \nomod{} observations suggest that, if recovery fails for a given secret $\textbf s$, the failure may be due to a low \nomod{} factor in the preprocessed training set. This reveals a strategy for potentially recovering such secrets. If an initial run of \system{} fails for a secret $\bf s$, start over with the (un-preprocessed) matrices $\mathbf A_i$, sampled from the original \lwe{} samples. For each of these, apply a random permutation $\Pi$ to the columns and preprocess the permuted matrices. This creates a new dataset of $\mathbf A'$ with corresponding $b'$, associated to a permuted counterpart $\mathbf s'$ of the original secret $\mathbf s$. If the \nomod{} of $\textbf s'$, $b'$ and $\mathbf A'$ is higher that that of $\mathbf s$, $b$ and $\textbf A$ (though this cannot be measured), $\textbf s'$ can be recovered. If the attack succeeds, $\mathbf s$ can be restored by applying the inverse of permutation $\Pi$. Table~\ref{tab:permute_cols} presents the impact of permutations on \nomod{} for $10$ secrets and $h=6,7,8$ for $n=256$ and $\log_2 q=14$. Some secrets become recoverable after using the permutation trick. 

\section{A Theoretical Scaling Law for \system{}}\label{sec:scaling_law}

\textbf{Intuition.} The \nomod{} experiments provide a key insight about the hardness of secret recovery in \salsa{}-based attacks on \lwe{}. They suggest that a secret $\mathbf s$ can be recovered from a training set if over $67\%$ of the $\{ x = \mathbf a \cdot \mathbf s - b \}$ are concentrated in the interval of length $q$. If the random variable $x$ is Gaussian (a reasonable assumption for $h\gg 1$, since the entries of $\mathbf a$ are random and bounded), $68\%$ of its values will be within one standard deviation of its mean, i.e. spread over two standard deviations. Therefore, for the secret to be recoverable, the standard deviation of $x$ should satisfy $\sigma_x \leq q/2$.
If $\mathbf s$ is a binary secret with Hamming weight $h$, and the entries of $\mathbf a$ have standard deviation $\sigma_a$, we have $x \approx \mathbf a \cdot \mathbf s - e$ and $\sigma_x \approx \sqrt{h}\sigma_a + \sigma_e \approx \sqrt{h}\sigma_a$.
Therefore, $\mathbf s$ is recoverable if $\sqrt{h}\sigma_a \leq q/2$, or $\sigma_a \leq \frac{q}{2\sqrt{h}}$.

\textbf{Scaling Laws.} We now apply this insight to the experimental results of ML-based attacks on \lwe{}. Consider the original \salsa{} attack, which does not utilize data preprocessing. Since the entries of $\mathbf a$ are uniformly distributed over $[0,q)$, $\sigma_a = \frac{q}{\sqrt{12}}$. 
Replacing in $\sigma_a \leq \frac{q}{2\sqrt{h}}$ yields $h \leq 3$, the main experimental result of \salsa{}. If we constrain the entries of $\mathbf a$ to be in $[0,\alpha q)$ (Table 4 in \cite{wengersalsa}), we have $\sigma_a = \frac{\alpha q}{\sqrt{12}}$, and $h \leq \frac{3}{\alpha^2}$. Applying this formula for $\alpha \in {0.6, 0.55, 0.5, 0.45}$, we obtain maximal recoverable Hamming weights of 8, 10, 12 and 15, which closely match \salsa{} experimental results.

These results shed light on the role of preprocessing in \picante{} and \system{}. When the standard deviation of $\mathbf a$ is reduced by a factor $\alpha$, maximal recoverable $h$ increases by a factor $\frac{1}{\alpha^2}$. However, the formula $h \leq \frac{3}{\alpha^2}$ underestimates actually recovered $h$ by a factor of $2$. For instance, from the reduction factors from Table~\ref{tab:512_results}, we should expect recovered $h$ to range from $11$ to $29$ as preprocessing time increases from 7 to 35 days, but actual recovered $h$ ranges from $20$ to $63$. As seen in \S\ref{sec:smaller_q}, the preprocessing step makes some secrets easier to recover for the same $h$, so \system{} performs better than predicted by theory for some secrets (at the expense of other secrets). Finally, note that the formula for the standard deviation of $x$ is the same for ternary and binary secrets. This accounts for the observation in \S\ref{sec:secret_dist} that ternary secrets are of similar difficulty for \system{} as binary secrets.

\section{Model architecture}
\label{sec:model}

\system{}'s baseline model uses a seq2seq architecture as in \salsa{} and \picante{}. Here, we compare it to the new encoder-only model discussed in \S\ref{sec:overview}. Specifically, this new model is based on the DeBERTa~\cite{deberta21} model 
but replaces its ``relative positional embeddings'' with rotary word embeddings, in the spirit of rotary positional embeddings (RoPE)~\cite{rope21} but applied to the integer words. It has 4 layers, 4 heads and 512 dimensions.
On top of the cross-entropy loss, we add an auxiliary, squared earth mover's distance (EMD) loss to compare model's softmax distribution with the target $b$. This encourages the model to make predictions that are at least close to the targets, if not exact matches. 
Trained with the auxiliary EMD loss, the model also replaces the beam search used in the distinguisher with a novel EMD-based distinguisher that compares the difference between the \emph{distributions} produced by the model on $\bf a_{\text{test}}$ and $\mathbf a_{\text{test}} + K\mathbf e_i$.

\begin{table}[h!]
\small
  \centering
  \vspace{-0.3cm}
\small\caption{\textbf{Performances of seq2seq and encoder-only models.} $n=256, \log_2q=12$; $n=512, \log_2q=41$. For binary and ternary secrets, we run 10 secrets per $h$ and indicate the epochs of full recovery.}
\vspace{1mm}
\label{tab:model_comparison}
\begin{tabular}{lccccccccccc}
\toprule
$n$, secret $\chi_s$  & \multicolumn{2}{c}{\bf 256, binary} \hspace{3mm} & \multicolumn{3}{c}{\bf 256, ternary} \hspace{3mm} & \multicolumn{3}{c}{\bf 512, binary} \hspace{3mm} & \multicolumn{3}{c}{\bf 512, ternary} \\ \cmidrule(lr){2-12}
$h$         & 5       & 6   \hspace{3mm}    & 4           & 5         & 6  \hspace{3mm}  & 57   \hspace{-2mm}     & 59   \hspace{-2mm}    & 63    \hspace{3mm}   & 55   \hspace{-2mm}     & 58   \hspace{-2mm}     & 60       \\ \midrule
Seq2seq       & 7        & 4,7  \hspace{3mm}    & 0,5,7,18     & 0,0,1      & 1 \hspace{3mm}   & 4    \hspace{-2mm}     & 8    \hspace{-2mm}    & 7     \hspace{3mm}   & 16   \hspace{-2mm}     & 11  \hspace{-2mm}      & -        \\
Encoder-only  & 20,23,27 & 16,23,28 \hspace{3mm} & 0,16,27      & 1,1,2,6    & 2   \hspace{3mm} & 2    \hspace{-2mm}     & 3   \hspace{-2mm}     & 3   \hspace{3mm}     & -    \hspace{-2mm}     & 6   \hspace{-2mm}      & 5    \\ \bottomrule
\end{tabular}
\vspace{-0.1cm}
\end{table}

Overall, we find that the performance of the two models are comparable (see Table~\ref{tab:model_comparison}). The encoder-only model requires more training epochs before recovery for $n=256, \log_2 q=12$, but requires fewer epochs for $n=512$, and may scale well to larger $n$. Furthermore, we observe that the EMD distinguisher still enables full recovery of ternary secrets with high $h$.

\section{Related Work}
\label{sec:back}

\textbf{ML for cryptanalysis.} Numerous proposals leverage ML for cryptanalysis, either indirectly or directly. We call approaches which use ML as part of (but not the main element of) the cryptanalysis process {\em indirect} approaches. Indirect approaches typically use ML models to strengthen existing cryptanalysis approaches, such as side channel or differential analysis~\cite{chen2021bridging}. Most relevant to this work is a recent paper showing the successful use of ML algorithms in side-channel analysis to attack Kyber, one of the NIST standardized PQC proposals~\cite{dubrova2022breaking}. {\em Direct} ML-based cryptanalysis schemes train models to directly recover cryptographic secrets from plaintext/ciphertext pairs or similar information. Such approaches have been studied against a variety of cryptosystems, including block ciphers~\cite{gohr2019improving, benamira2021deeper, chen2021bridging, alani2012neuro, so2020deep, kimura2021output, baek2020recent}, hash functions~\cite{goncharov2019using}, and substitution ciphers~\cite{ahmadzadeh2021novel, srivastava2018learning, aldarrab2020can, greydanus2017learning}. The two ML-based LWE attacks described in \S\ref{sec:intro}, \salsa{}~\cite{wengersalsa} and \picante{}~\cite{li2023salsa}, fall under this heading.

\textbf{Use of transformers for mathematics.} In recent years, language models have been used to solve math problems beyond the cryptanalysis applications explored in this work. Much prior work has considered how well language models can solve written math problems~\cite{hochreiter1997long, saxton2019analysing}. More recently, \cite{griffith021} showed large transformers could achieve high accuracy on elementary/high school problems, and~\cite{zong2022solving} explored the accuracy of GPT-3~\cite{brown2020language} on math word problems. Language models have also been applied to formalized symbolic math problems. After~\cite{lample2019deep} demonstrated that transformers can solve such problems with high accuracy, follow-up work has explored transformers' use in theorem proving \cite{polu2020generative}, dynamical systems \cite{charton2020learning}, SAT solving \cite{shi2021transformerbased}, transport graphs \cite{charton2021deep}, and symbolic regression \cite{biggio2021neural, dascoli2022deep}. Finally, some have proposed customizing model architectures to enable specific arithmetic operations~\cite{kaiser2015neural, trask2018neural, power2022grokking}.

\section{Discussion}
\label{sec:discuss}

We present \system{}, a ML-based attack on \lwe{} with sparse small secrets. \system{} improves data preprocessing and secret recovery, enabling significant performance gains over prior work, \salsa{}~\cite{wengersalsa} and \picante~\cite{li2023salsa}. In particular, \system{} can recover secrets with more general distributions (ternary, small Gaussian), larger dimensions $n$, smaller moduli $q$, and higher $h$. In our implementation, \system{} outperforms the uSVP attack, requiring less time but more compute resources (details in Appendix~\ref{sec:Compare_uSVP}). Most importantly, this work provides key theoretical insights into observed attack performance, paving the way for targeted future work. 
Note that even if we recover secrets with seemingly low probability, such as one seed out of ten succeeds for that secret ($1/10$),  or if we do not recover all secrets successfully, such as  we recover one out of ten secrets with our attack, this is still enough to make these cryptosystems unsafe to use with low weight secrets (unless there is a way to check for vulnerability to our attack without running the attack). 

\textbf{Limitations and broader impact.} Despite significantly advancing the state-of-the-art in ML-based \lwe{} attacks, \system{} cannot yet break standardized \lwe{}-based PQC schemes, limiting its real-world impact. Because of this, our paper raises no immediate security concerns. Nevertheless, we have shared a copy of our paper with the NIST PQC group to make them aware of this attack. 

\textbf{Future work.} Scaling \system{} to attack real-world systems requires work in several directions: increasing dimension $n$, reducing modulus $q$, and increasing $h$, the number of nonzero entries in recoverable secrets. 
Continued innovations in model architecture (a la \S\ref{sec:model}) may allow attacks in higher dimension $n$, while the theoretical scaling law of \S\ref{sec:scaling_law} provides helpful guidance for improving $q$ and $h$. 
Given our new insight and analysis of the importance of the \nomod{} percentage, we conclude that the reason that small $q$ is hard for the transformers with our current approach is that they are not good at modular arithmetic (not yet, anyway\textemdash this is an area for future work and improvement). In addition,  our preprocessing is not well-suited to what is precisely needed and relies on existing lattice reduction algorithms which scale poorly with $n$ and $q$.  

This suggests two avenues for improvement: first, to develop model architectures that perform modular arithmetic better. Limited existing work has studied the application of ML to modular arithmetic more broadly~\cite{power2022grokking, gromov2023grokking}. Second, alternative preprocessing techniques should be developed to directly concentrate the distribution of random vectors, without relying on lattice reduction methods. With the goal of reducing the standard deviation of the training data, around any center, techniques from the broader math community may prove helpful. 

\para{Acknowledgements.} We thank Jana Sot\'{a}kov\'{a} for her contributions to developing the attack on ternary secrets, Hamming reduction techniques, and Section A.3.
We also thank Mark Tygert for his helpful input.

\newpage

\bibliographystyle{plain}
\bibliography{references}


\newpage

\appendix

\section{Appendix}

\subsection{Parameters}
\label{subsec:params}
\begin{table}[h!]
\vspace{-4mm}
\small
  \centering
    \small\caption{\textbf{\lwe{}, preprocessing, and training parameters}. For the adaptive increase of preprocessing parameters, we start with blocksize $\beta_1$ and LLL-delta $\delta_{LLL1}$, and upgrade to $\beta_2$ and $\delta_{LLL2}$ at a later stage. Parameters base $B$ and bucket size $r$ are used to tokenize the numbers for transformer training. }
    \label{tab:params}
      \vspace{2mm}
\begin{tabular}{lllllllll}
\toprule
$n$                    & $\log_2 q$ & $q$             & $\beta_1$ & $\delta_{LLL1}$ & $\beta_2$ & $\delta_{LLL2}$ & base $B$         &  bucket size $r$         \\ \midrule
 \multirow{5}{*}{256} & 12   & 3329          & 35   & 0.99  & 40   & 1     & 417          & 1         \\
                     & 14   & 11197         & 35   & 0.99  & 40   & 1     & 1400         & 1         \\
                     & 16   & 42899         & 35   & 0.99  & 40   & 1     & 5363         & 4         \\
                     & 18   & 222553        & 35   & 0.99  & 40   & 0.99  & 27820        & 16        \\
                     & 20   & 842779        & 35   & 0.99  & 40   & 0.99  & 105348       & 64        \\ \midrule
\multirow{2}{*}{350} & 21   & 1489513       & 30   & 0.96  & 40   & 0.99  & 186190       & 128       \\
                     & 27   & 94056013      & 30   & 0.96  & 40   & 0.99  & 5878501      & 4096      \\ \midrule
512                  & 41   & 2199023255531 & 18   & 0.93  & 22   & 0.96  & 137438953471 & 134217728 \\ \bottomrule
\end{tabular}
\end{table}

\vspace{-2mm}
\system{} runs data preprocessing with the parameters shown in Table~\ref{tab:params} in parallel using multiple CPUs. We fully parallelize when the time to process one matrix is greater than $24$ hours\textemdash e.g., for $n=350, \log_2 q=27$, we used 5000 CPUs. Otherwise, we parallelize on fewer CPUs depending on the attack time allowed\textemdash e.g., preprocessing is completed on 270 CPUs in less than 4 days for $n=256, \log_2 q=14$, and on 990 CPUs in less than 3 days for $n=256, \log_2 q=18$. 

In \system{}, the tokenization used by the transformer mirrors the strategy in \S4 of~\cite{li2023salsa}, but uses smaller bases $B$ and larger bucket sizes $r$ for better performance (Table 9, 10 of~\cite{li2023salsa}). Decreasing $B$ is further supported by Table ~\ref{tab:base}, evaluated on a low modulus data using \system{}'s preprocessing ($n=256, \log_2q=16$). Full ternary secret recovery and partial Gaussian secret recovery both improve with smaller $B = 5363$. 

\begin{table}[h!]
\small
  \centering
  \vspace{-0.4cm}
      \small\caption{\textbf{Secret recovery for different bases}. $n=256, \log_2q=16$. We show full ternary secret recovery and partial Gaussian secret recovery when using $B = 7150$ and $5363$ on the same datasets and secrets. }
    \label{tab:base}
    \vspace{2mm}
\begin{tabular}{lccccccccc}
\toprule
         & \multicolumn{5}{c}{ternary}      & \multicolumn{4}{c}{Gaussian} \\
$h$     \hspace{5mm}   & 8    & 9    & 10   & 11   & 12  \hspace{5mm} & 3     & 4     & 5     & 6    \\ \midrule
$B = 7150$ \hspace{5mm} & 2/10 & 0/10 & 1/10 & 0/10 & 1/10\hspace{5mm} & 5/10  & 6/10  & 1/10  & 1/10 \\
$B = 5363$ \hspace{5mm} & 2/10 & 0/10 & 2/10 & 0/10 & 2/10 \hspace{5mm} & 5/10  & 6/10  & 1/10  & 3/10 \\ \bottomrule
\end{tabular}

\end{table}

\subsection{More seeds for initialization}
\label{subsec:seeds}
The ML attacks benefit from running multiple times with different seeds, or {\it initializations}, as was demonstrated in \picante{}~\cite[Section 6.4]{li2023salsa}. More seeds improve both the success probability and the number of epochs required. 
Table~\ref{tab:seed} shows how binary secret recovery improves with more seeds, for different Hamming weights $h$ when $n=256$ and $\log_2 q = 12$. 

\begin{table}[h]
\vspace{-4mm}
\small
  \centering
    \small\captionof{table}{\textbf{Secret recovery with 1 vs 5 seeds}. $n=256$, $\log_2 q=12$, binary secrets. For 1 seed, $epoch$ is the epoch of secret recovery; For 5 seeds (ran on the same secrets as the 1 seed experiments), $epoch$ is the lowest epoch of secret recovery among the 5 initializations for each secret. }
    \label{tab:seed}
    \vspace{2mm}
\begin{tabular}{lllll}
\toprule
$h$                 & 3              & 4     & 5    & 6    \\ \midrule
recovery, 1 seed & 5/10           & 1/10  & 1/10 & 1/10 \\
$epoch$             & 0,0,0,8,17     & 0     & 17   & 13   \\ \midrule
recovery, 5 seeds & 7/10           & 3/10  & 1/10 & 2/10 \\
$epoch$             & 0,0,0,7,7,8,17 & 0,5,5 & 7    & 4,7  \\ \bottomrule
\end{tabular}
\end{table}

Table~\ref{tab:512_seeds} shows the results for ternary secrets on $n=512$, $\log_2 q = 41$, where we run 5 initializations for each secret. While most initializations partially recovered the secret, only a few got full recovery within 20 epochs. Full recovery benefits from more initializations, especially for high $h$.

\begin{table}[h!]
\vspace{-4mm}
\small
  \centering
    \small\caption{\textbf{Ternary secret recovery with 5 initializations}. $n=512$. `-': recovery did not occur in $\le 20$ epochs. }
    \label{tab:512_seeds}
      \vspace{2mm}
\begin{tabular}{lllllll}
\toprule
$h$                & 45        & 46        & 50         & 55        & 58       & 60         \\ \midrule
epoch of partial recovery          & 4,4,6,6,7 & 1,1,1,1,1 & 3,4,5,7,-    & 5,6,6,6,7 & 6,6,8,10 & 5,8,8,8,12 \\
epoch of full recovery          & 10,-,-,-,-        & 5,7,-,-,-       & 4,10,14,17,- & 16,-,-,-,-        & 11,11,-,-,-    &           -,-,-,-,- \\ \bottomrule
\end{tabular}
\end{table}

\subsection{Comparison with \picante{} }
\label{sec:picante}

To demonstrate the power of the new preprocessing in \salsys{}, we run a set of experiments on $n=256$, $\log_2 q = 23$, where \picante{} also showed success. Using blocksize $\beta=40$, each matrix is processed by \system{} in about $1.5$ days; \picante{} took $2.2$ days/matrix. \system{} achieves a reduction factor of $0.25$, compared to $0.33$ in \picante{}. As shown in Table~\ref{tab:highest_h_256}, the difference in the preprocessing step is even more striking for lower $q$.

The highest $h$ recovered by \picante{} was $h=31$ in 4 out of 20 experiments; \system{} recovered $h=43$. 
In Table~\ref{table:verde_picante_epochs}, we see that for $h=26-31$, \system{} significantly outperforms \picante{} in both the success rate and the number of epochs required.  In other words, better preprocessing results in lower training time and better secret recovery.
\begin{table}[h]
\vspace{-0.3cm}
\caption{{\bf Epochs of secret recovery for \picante{} vs. \system{}.} $n=256$ and $\log_2 q = 23$. `-' means secret not recovered. 5 secrets per $h$, except for \picante{} $h=31$ (20 secrets).}
\label{table:verde_picante_epochs}
\vspace{2mm}
\small
\centering
\begin{tabular}{lllllll}
\toprule
$h$ & 26 & 27 & 28 & 29 & 30      & 31           \\ \midrule
\picante{}                                                            &  2,3,4,7,- & 10,-,-,-,- & 5,-,-,-,- & 5,9,11,-,- & 17,20,32,-,- & 6,12,26,27 (out of 20 secrets)\\
\system{} & 0,0,2,2,7 & 2,6,-,-,- & 0,0,0,1,2 & 0,1,1,1,- &  0,1,2,3,- & 1,2,3,4,-\\ \bottomrule
\end{tabular}

\vspace{-0.5cm}
\end{table}

\subsection{Distinguisher tested on reduced data outperforms random data}
\label{subsec:dist_bkz}
We compare the performance of running the distinguisher on the preprocessed data that were held out from the training set ($\text{Dist}_{\text{BKZ}}$) with running on random vectors ($\text{Dist}_{\text{rand}}$). We run both set of experiments with the same model initialization seeds, and record a success for the method(s) that recovers the secret at the earliest epoch. Table~\ref{tab:distinguisher_bkz} indicates that  $\text{Dist}_{\text{BKZ}}$ performs better. 

\begin{table}[h!]
\small
  \centering
\vspace{-0.4cm}
    \caption{\textbf{Secret recovery by running the distinguisher on the random vectors and bkz preprocessed data}. $n=256, \log_2 q=23$, on data processed using \picante{}'s approach. }
  \label{tab:distinguisher_bkz}
    \vspace{2mm}
\begin{tabular}{llllll}
\toprule
$h$  & 27 & 28 & 29 & 30 & 31 \\ \midrule
$\text{Dist}_{\text{BKZ}}$ & 2/5  & 1/5  & 1/5  & 0/5  & 0/5  \\
$\text{Dist}_{\text{rand}}$ & 1/5  & 0/5  & 0/5  & 0/5  & 0/5  \\ 
\bottomrule
\end{tabular}
 \vspace{-0.3cm}
\end{table}

\subsection{Dimension reduction techniques}
\label{sec:reduction}

Most of the entries in a sparse secret are zero, so prior work~\cite{Albrecht2017_sparse_binary} has suggested the idea of randomly assuming a subset of the entries to be zero and removing them to reduce the dimension of the lattice problem.  The assumption will be correct with some probability depending on the secret's sparsity.   
Here we explore an improvement on this strategy: we use the partially trained model to glean signal on which entries should be kicked out.
We can either try to kick out zeros, which we call {\it dimension reduction}, or in the binary case, kick out $1$s, which we call {\it Hamming reduction}, or combined. 

This technique will be better than random when the model has begun to learn information about the bits, reflected in their relative rankings.  Specifically, the ranking strategies described in
~\cite[Section 4.3]{li2023salsa} are used to compute scores which estimate the likelihood of secret bits being $1$. 
Once the model has started to learn, we can assume that 
the highest ranked bits will correspond to secret bits which are equal to $1$, and the lowest ranked bits will correspond to zeros.  
So we use this information to reduce the dimension of the problem by kicking out low-ranked bits which we guess to be zero or high-ranked bits which we guess to be $1$.
Then, we retrain a model on the smaller dimensional samples and hope to recover the secret. If the original kicked out bits were correct and the model recovers the secret of the smaller dimensional problem, then we find the original secret. 

{\bf Dimension reduction.} Since there are many more $0$s than $1$s in sparse secrets, we can potentially reduce the dimension significantly. Once we remove the bits with low scores, we can simply re-run training on the dataset with $(\textbf{a}', b)$ where $\textbf{a}'$ are the samples with the corresponding bits removed. If the indices have been identified incorrectly, then the reduction will fail. For $n=256, \log_2q=14$, \system{}  attempted 10 binary secrets with $h=10$ and did not recover the secret. Then we tried dimension reduction on these experiments and recovered one secret. 

{\bf Hamming reduction.} Kicking out $1s$ from the secret is particularly valuable, given the theoretical analysis of the \system{} in Section~\ref{sec:scaling_law}. 
If nonzero bits are indeed ranked at the top by the model, a straightforward approach of kicking out the top-ranked bits and retraining on the smaller dimension and Hamming weight will likely
yield improved secret recovery.

But in case some of the top-ranked bits are not equal to $1$, we propose the following strategy.
Suppose $S$ is a small set of indices for bits with the highest scores. 
We construct the following problem: let $\textbf{s}'$ be $\textbf{s}$ with bits in $S$ flipped, and $\textbf{a}'$ be $\textbf{a}$ with $a_i$ negated for $i\in S$. Equivalently, for $i\in S$, $a'_i=-a_i$ and $s'_i=1-s_i$. Then, the corresponding 
$$b' = \mathbf{a' \cdot s'} = \sum_{i\not\in S} a_i s_i + \sum_{i\in S} a'_i s'_i = \sum_{i\not\in S} a_i s_i + \sum_{i\in S} -a_i (1-s_i) = b - \sum_{i \in S} a_i.$$ 
If more than half of the indices in $S$ are $1$, then $\textbf{s}'$ has a smaller Hamming weight, hence the instance $(\textbf{a}',b'=b - \sum_{i \in S} a_i)$ is likely easier. If exactly half of the indices in $S$ are $1$, then $\textbf{s}'$ has the same $h$ as $\textbf{s}$, but the new instance $(\textbf{a}',b')$ will have a different \nomod{} and may be recoverable.  

\subsection{Attacking sparse secrets in larger dimensions}
\label{subsec:large_dim}
For sparse secrets on even larger dimensions, we can apply our attack after using combinatorial techniques to exploit the sparsity of the secret. The approach would be to combine \system{} with the techniques from~\cite{Albrecht2017_sparse_binary, Cheon_hybrid_dual} as follows: Randomly kick out $k$ entries of the secret, assuming they are zero, which will be true with some probability. 
This reduces the \lwe{} problem to a smaller dimension where \system{} can recover the secret. The expected cost of the attack would be \system{}'s cost multiplied by $1/p$, where $p=(\frac{n-h}{n})^k$ is the probability that the assumption that the $k$ entries are $0$ is correct. 

\subsection{Comparison with lattice reduction/uSVP attacks}
\label{sec:Compare_uSVP}

In this section we compare \system{} with classical lattice reduction attacks in two ways.  The \lwe{} Estimator \cite{LWEestimator} gives heuristic predicted running times for the best-known lattice reduction attacks.  But even the authors of the \lwe{} Estimator claim that the estimates are often wrong and unreliable. So we compare \system{} to the Estimator results but we also compare to concrete running times in the case $n=256$, achieved by implementing the uSVP lattice attacks ourselves, on the same machines where we run our ML-based attacks. Unfortunately, many attacks listed by the \lwe{} Estimator lack practical/accessible implementations, and we lacked the time and resources to implement all of these and run comparisons. Thus, we focus our concrete comparisons on the uSVP attack, which the Estimator predicts to be the best method for larger $n$ such as $n=512$.  We leave comparison of \system{} against other lattice reduction attacks as important future work for the broader lattice community, as practical attack run-times are poorly understood, especially for small sparse secrets.

\begin{table}[h]
\vspace{-4mm}
\small
\centering

\caption{{\bf Concrete comparison of \system{} and uSVP attacks for $n=256$, binary secrets, varying $q$ and $h$.} \system{}'s total attack time is the sum of preprocessing and training time (with recovery included). Preprocessing time assumes full parallelization, and training time is the number of epochs to recovery multiplied by epoch time ($1.5$ hours/epoch). {\bf fail} means no successful secret recovery for uSVP to compare to.
}

\label{tab:binary_verde_usvp}
\vspace{2mm}
\begin{tabular}{cccccc}
\toprule
\multicolumn{2}{c}{\textbf{\lwe{} parameters}} & \multicolumn{3}{c}{\textbf{\system{} attack time}} & \multirow{2}{*}{\textbf{uSVP attack time (hrs)}} \\
$\log_2 q$ & $h$ & \textit{Preprocessing (hrs)} & \textit{Training} & \textit{Total (hrs)} &  \\ \midrule
12 & 8 & 1.5 & 2 epochs & 4.5 &  {\bf fail} \\
14 & 12 & 2.5 & 2-5 epochs & 5.5-10 &  {\bf fail} \\
16 & 14 & 8.0 & 2 epochs & 11 & {\bf fail} \\
18 & 18 & 7.0 & 3 epochs & 11.5 & 558 \\
18 & 20 & 7.0 & 1-8 epochs & 8.5-19 & 259 \\
20 & 22 & 7.5 & 5 epochs & 15 & 135-459 \\
20 & 23 & 7.5 & 3-4 epochs & 12-15 & 167-330 \\
20 & 24 & 7.5 & 4 epochs & 13.5 & 567 \\
20 & 25 & 7.5 & 5 epochs & 15 & 76 - 401 \\ \bottomrule
\end{tabular}

\end{table}

Table~\ref{tab:binary_verde_usvp} presents our concrete comparison between \system{} and the uSVP attack on binary secrets for $n=256$.  To summarize the comparison, 
\system{} outperforms the uSVP attack in two senses: 
1) \system{} fully recovers sparse binary and ternary secrets for $n$ and $q$ in some cases where the uSVP attack does not succeed in several weeks or months using \fplll{} BKZ 2.0 ~\cite{CN11_BKZ} with the required block size; and 2) in cases where we improve the uSVP implementation enough (see below) to run with the required large block size, we find that in all cases, \system{} recovers the secrets much faster.

{\bf Summarizing classical lattice reduction attacks.}
Table~\ref{tab:estimator} gives the estimated heuristic cost and specifies the block size for attacking sparse binary and ternary secrets for $n = 256, 350, 512$ and various $q$ with the best known classical lattice reduction attack.  

\begin{table}[h!]
\small
  \centering
      \small\caption{\textbf{Estimated cost of best classical attack (\lwe{} Estimator)}. For \system{}'s highest $h$, we run the \lwe{} Estimator and report the estimated cost and block size $\beta$ for the best predicted attack. }
    \label{tab:estimator}
    \vspace{2mm}
    \centering
\begin{tabular}{llllllllll}
\toprule
\multirow{2}{*}{$n$} & \multirow{2}{*}{$\log_2 q$} & \multicolumn{4}{c}{\bf binary secret}    \hspace{5mm}                    & \multicolumn{4}{c}{\bf ternary secret}    \\  \cmidrule{3-10}
                   &                         & h  & best attack                           & rop  & $\beta$   \hspace{5mm}   & h  & best attack        & rop  & $\beta$ \\ \midrule
\multirow{5}{*}{256}                & 12                      & 8  & dual\_mitm\_hybrid                          & $2^{43.0}$ & 40    \hspace{5mm}   & 9  & dual\_mitm\_hybrid & $2^{43.8}$ & 40   \\
                & 14                      & 12  & dual\_hybrid                          & $2^{47.2}$ & 40   \hspace{5mm}    & 13  & dual\_hybrid  & $2^{47.7}$ & 41   \\
                & 16                      & 14 & dual\_hybrid                    & $2^{46.8}$ & 40   \hspace{5mm}    & 16 & dual\_hybrid & $2^{47.4}$ & 41   \\
               & 18                      & 23 &    bdd\_hybrid                      & $2^{47.1}$ &  45  \hspace{5mm}    & 23 &   dual\_hybrid     & $2^{47.4}$ &   41 \\
               & 20                      & 36 &     bdd                             & $2^{44.0}$ &  45   \hspace{5mm}   & 33 &       bdd         & $2^{44.2}$ & 46   \\ \midrule
\multirow{2}{*}{350}                & 21                      & 12 & \begin{tabular}[c]{@{}l@{}}dual\_mitm\_hybrid, \\ bdd\_mitm\_hybrid\end{tabular} & $2^{46.4}$ & 40   \hspace{5mm}    & 13 & dual\_mitm\_hybrid & $2^{46.9}$ & 54   \\
                & 27                      & 36 & bdd                                   & $2^{44.9}$ & 47   \hspace{5mm}    & 38 & bdd, bdd\_hybrid   & $2^{44.1}$ & 43   \\ \midrule
512                & 41                      & 63 & usvp/bdd                              & $2^{42.9}$ & 40   \hspace{5mm}    & 60 & usvp/bdd           & $2^{42.9}$ & 40  \\ \bottomrule
\end{tabular}
\end{table}

\begin{table}[h!]
\vspace{-0.3cm}
\small
  \centering

    \small\caption{\textbf{Estimated cost and block sizes of uSVP}. $n=256, \log_2q = 16,18,20$, for binary, ternary, and Gaussian secrets with $h$ nonzero entries. }
    \label{tab:estimator_usvp}
      \vspace{2mm}
\begin{tabular}{lccccccccc}
\toprule
\multirow{2}{*}{$\log_2 q$} & \multicolumn{3}{c}{\bf binary} \hspace{5mm} & \multicolumn{3}{c}{\bf ternary} \hspace{5mm} & \multicolumn{3}{c}{\bf Gaussian} \\
               &   $h$     & rop          & $\beta$     \hspace{5mm}    
               &  $h$ & rop          & $\beta$    \hspace{5mm}      
               & $h$ &  rop      & $\beta$         \\ \midrule
16      &   12            & $2^{54.7}$         & 86     \hspace{5mm}      
 &   12 & $2^{54.9}$         & 87    \hspace{5mm}        
 &   6 & $2^{69.3}$          & 138          \\
18        &18             & $2^{49.5}$         & 67     \hspace{5mm}   
 & 19 & $2^{49.8}$         & 68      \hspace{5mm}   
 & 7 & $2^{59.5}$          & 102          \\
20            & 25         & $2^{45.4}$         & 52   \hspace{5mm}   
& 24 & $2^{45.4}$         & 52    \hspace{5mm}      
& 7 & $2^{53.5}$          & 80 \\ \bottomrule          
\end{tabular}
\vspace{0.3cm}
\end{table}

Table~\ref{tab:estimator_usvp} gives the same information for the uSVP classical lattice reduction attack, focusing on $n=256$ for some of the larger $q$ and $h$ where \system{} succeeds.

{\bf uSVP attack performance, binary secrets.} For $n=256$ and $\log_2 q = 20$, we run the concrete uSVP attack using \fplll{} BKZ 2.0 with Kannan's embedding and parameters (\cite{Kan87}, \cite{CCLS}). 
With block size $50$ and $55$, secrets are not recovered in $25$ days, and block size $60$ or larger cannot finish the first BKZ loop in $3$ days. 
We propose two improvements to get these attacks to run faster and better: 1) we rearrange the rows of the uSVP matrix as in \system{};  2) we use the adaptive float type upgrade as in \system{}. In addition, after each BKZ loop, we also run the secret validation obtained from the shortest vector found so far, and terminate if we get a secret match. 

With rearranging the rows but without the adaptive float type upgrade, we have to use higher precision because otherwise the attack fails due to low precision after running for $23$ hours. The attacks run quite slowly with high precision and did not recover secrets after $25$ days, except in one case where a binary secret with $h=22$ was recovered with block size $55$ in $414$ hours, roughly $17$ days. 

\begin{table}[h!]
\small
  \centering

  \vspace{-4mm}
    \small\caption{\textbf{uSVP concrete attack time, $n=256$.} Upper: $\log_2 q=20$, lower: $\log_2 q=18$. For each $h$, block size and secret distribution, we run 5 experiments on different secrets and show the secret recovery time (in hours). `-' means no success in 25 days. }
    \label{tab:concrete_usvp}
    
$\bf n=256, \log_2q=20$
\begin{tabular}{llllllll}
\toprule
blocksize           & secret  & $h=22$          & $h=23$          & $h=24$          & $h=25$         &    &     \\ \midrule
\multirow{2}{*}{50} & binary  & -             & -             & -             & 451, 531     &    &     \\
  & ternary & 261, 367      & -             & -             & -            &    &     \\ \midrule
\multirow{2}{*}{55} & binary  & 135, 161, 459 & 167, 323, 330 & 567           & 76, 280, 401 &    &     \\
  & ternary & 43, 241, 432  & 234, 296      & 226, 257, 337 & 280          &    &     \\ \bottomrule
\end{tabular}

\vspace{2mm}
$\bf n=256, \log_2q=18$
\begin{tabular}{llllllll}
\toprule
blocksize           & secret   & $h=11$  & $h=12$  & $h=17$  & $h=18$  & $h=19$  & $h=20$  \\ \midrule
\multirow{2}{*}{65} & binary  & 32            & -             & -             & 558          & -  & 259 \\
 & ternary & 324, 553      & 109           & 594           & -            & -  & -  \\ \bottomrule
\end{tabular}
\vspace{-0.2cm}
\end{table}

With rearranging the rows and the adaptive float type upgrade, we ran $n=256$ and $\log_2 q = 20$ with block size $50-55$ for binary/ternary secrets, and $n=256$ and $\log_2 q = 18$ with block size $65$ for binary/ternary secrets. 
See Table~\ref{tab:concrete_usvp} for running times and $h$.  For example, for $\log_2 q = 20$, a ternary secret with $h=25$ was recovered in $280$ hours, roughly $12$ days, and for $\log_2 q = 18$, a binary secret with $h=20$ was found in $\approx 11$ days. When the block size used is lower than predicted by the estimator (50 instead of 52 for $\log_2 q = 20$ and 65 instead of 67 for $\log_2 q = 18$, see Table~\ref{tab:estimator_usvp},~\ref{tab:concrete_usvp}), the uSVP attack succeeds in recovering only a few secrets out of many which were tried. 

The concrete experiments allow us to validate to some extent the predictions of the Estimator in many cases, giving confidence in the comparison with \system{}.  Running the uSVP attack with block size $70$ or larger didn't finish the first BKZ loop in $3$ days, so we expect longer attack time for Gaussian secrets and for binary and ternary secrets with lower $q$.

{\bf Gaussian secrets.} \system{} achieves partial Gaussian secret recovery for small $h$, reducing the secret recovery to a lattice problem in tiny dimension ($h$). Because the preprocessing time and per epoch time does not vary with secret distribution, \system{}'s attack time on Gaussian secrets is comparable to on binary secrets (see Table~\ref{tab:binary_verde_usvp}, \ref{tab:gaussian_verde}). We note that preprocessing takes longer for $n=350$ and $\log_2 q=27$ due to precision issues. This is an important observation because lattice problems are supposed to be harder for smaller $q$. In contrast, with classical attacks, the Estimator predicts significantly larger block sizes required, and longer running times (Table \ref{tab:estimator_g}), than for binary secrets. \system{} compares favorably to classical attacks on Gaussian secrets, especially on small $q$ (where the problem is harder), e.g., \system{} recovers Gaussian secrets with $h=5-6$ in $4.5-15$ hours for $n=256, \log_2 q=12$, where the cost of best classical attacks is predicted to be $2^{91}$ rop. 

\begin{table}[h]
\small
\centering
\vspace{-0.2cm}

\caption{{\bf \system{}'s performance on \lwe{} problems with $n=256$ and $350$, Gaussian secrets, varying $q$ and $h$.} Preprocessing time: hours to process one matrix. Total attack time: sum of preprocessing time (assuming full parallelization) and training time (number of epochs multiplied by hours per epoch, see \S\ref{sec:overview}).
}

\label{tab:gaussian_verde}
\vspace{2mm}
\begin{tabular}{ccccccc}
\toprule
\multicolumn{3}{c}{\textbf{\lwe{} parameters}} & \multicolumn{3}{c}{\textbf{\system{} attack time}} \\
$n$ & $\log_2 q$ & $h$ & \textit{Preprocessing (hrs)} & \textit{Training} & \textit{hrs/epoch} & \textit{Total (hrs)}  \\ \midrule
\multirow{5}{*}{256} & 12 & 5-6 & 1.5 & 2-9 epochs & \multirow{5}{*}{1.5} & 4.5-15  \\
 & 14 & 5-6 & 2.5 & 2-21 epochs & & 5.5-34  \\
 & 16 & 9 & 8.0 & 2 epochs & & 11 \\
 & 18 & 9 & 7.0 & 3 epochs & & 11.5  \\
 & 20 & 10 & 7.5 & 5 epochs & & 15 \\ \midrule
 \multirow{2}{*}{350} & 21 & 5 & 16 & 1-5 epochs & \multirow{2}{*}{1.6} & 18-24  \\
 & 27 & 10 & 216 & 2-13 epochs & & 219-237  \\ \bottomrule
\end{tabular}
\vspace{-0.2cm}
\end{table}
\begin{table}[h!]
\small
  \centering

    \small\caption{\textbf{\lwe{} Estimator: best classical attack on Gaussian secrets on various $q$.} $n=256$ and $350$. For \system{}'s highest $h$, we show the best classical attack, the attack cost (rop), and predicted block size $\beta$ from the \lwe{} Estimator.}
    \label{tab:estimator_g}
      \vspace{2mm}
\begin{tabular}{llllll}
\toprule
$n$                    & $\log_2 q$ & $h$  & best attack      & rop    & $\beta$ \\ \midrule
\multirow{5}{*}{256} & 12   & 6  & bdd / bdd\_hybrid & $2^{91.0}$ & 214  \\
                     & 14   & 6  & bdd / bdd\_hybrid & $2^{77.7}$ & 166  \\
                     & 16   & 9  & bdd / bdd\_hybrid & $2^{67.1}$ & 128  \\
                     & 18   & 9  & bdd            & $2^{57.7}$ & 93   \\
                     & 20   & 10  & bdd / bdd\_hybrid & $2^{51.9}$ & 72    \\ \midrule
\multirow{2}{*}{350} & 21   & 5  & bdd              & $2^{65.2}$ & 120  \\
                     & 27   & 10 & bdd / bdd\_hybrid & $2^{50.2}$ & 68   \\  
\bottomrule 
\end{tabular}
\end{table}

\end{document}